\documentclass[modern]{aastex63}

\begin{document}

\title{Thirty-Three New Stellar Angular Diameters from the NPOI, and Nearly 180 NPOI Diameters as an Ensemble}

\author{Ellyn K. Baines}
\affil{Naval Research Laboratory, Remote Sensing Division, 4555 Overlook Ave SW, Washington, DC 20375, USA}
\email{ellyn.baines@nrl.navy.mil}

\author{James H. Clark III}
\affil{Naval Research Laboratory, Remote Sensing Division, 4555 Overlook Ave SW, Washington, DC 20375, USA}

\author{Henrique R. Schmitt}
\affil{Naval Research Laboratory, Remote Sensing Division, 4555 Overlook Ave SW, Washington, DC 20375, USA}

\author{Jordan M. Stone}
\affil{Naval Research Laboratory, Remote Sensing Division, 4555 Overlook Ave SW, Washington, DC 20375, USA}

\author{Kaspar von Braun}
\affil{Lowell Observatory, 1400 W. Mars Hill Rd, Flagstaff, AZ 86001, USA} 

\begin{abstract}

We present new angular diameter measurements for 33 stars from the Navy Precision Optical Interferometer, reaching uncertainties on the limb-darkened diameter of 2$\%$ or less for 21 targets. We also determined the physical radius, bolometric flux, luminosity, and effective temperature for each star. Our sample is a mix of giant, subgiant, and dwarf stars, and span spectral classes from mid-A to to mid-K. We combined these 33 stars with samples from previous publications to analyze how the NPOI diameters compare to those obtained using other means, namely ($V-K$) color, the \emph{JMMC Stellar Diameters Catalog}, and $Gaia$ predictions.  

\end{abstract}

\keywords{stars: fundamental parameters, techniques: high angular resolution, techniques: interferometric}


\section{Introduction} 

In one sense, the story of astronomy can be told as the quest for better resolution: in its simplest form, the larger the telescope, the more detail you can see on a celestial object. At a certain point, extremely large telescope mirrors become incredibly complicated and prohibitively expensive to build, so we use telescope arrays to provide the ever-increasing resolution required. Optical and infrared interferometry has been used for some exciting explorations, including an expanding fireball from a nova explosion \citep{2014Natur.515..234S}, observations of the dust sublimation region of a bright AGN \citet{2022ApJ...940...28K}, determining the size and thermal properties of an asteroid \citep{2013Icar..226..419M}, supporting theoretical descriptions of a Mira variable star's atmosphere \citep{2016AandA...587A..12W}, and so on. Since its inception, optical and infrared interferometry has been used to measure stellar angular diameters \citep[][and many more]{1921ApJ....53..249M, 2001AandA...377..981W, 2009MNRAS.394.1925V, 2012ApJ...757..112B, 2017AandA...597A.137K}, though these measurements are more generally the exception rather than the rule.

Stellar diameters have historically been determined using indirect methods, with photometry and spectroscopy being the most common. However, both of these techniques rely upon models of stellar interiors and atmospheres that cannot fully describe the complexity of the stars themselves. To make the models feasible, a number of simplifications and/or assumptions are required that we hope are mostly right, but some evidence shows they are not always accurate \citep [e.g.,][showed that models overestimate cooler stellar temperatures by $\sim 3\%$ and underestimate radii by $\sim 5\%$ for small stars. This paper also includes a discussion about the discrepancy between model predictions and direct measurements]{2012ApJ...757..112B}. Interferometric measurements are key to testing stellar models and acting as benchmarks \citep[e.g.,][]{2020AandA...642A.101P, 2020AandA...640A..25K}, which is particularly important in high signal-to-noise stellar spectroscopic studies and the \emph{Gaia} survey. A collection of reliably calibrated stellar radii and effective temperatures based on accurate diameters is vital for their use in determining evolutionary state, understanding any planets orbiting the star, and calibrating empirical relationships such as the photometric color-temperature scale \citep{2020MNRAS.493.2377R}.

The angular diameter measurements presented here are a continuation of the survey project in \citet{2018AJ....155...30B} and \citet{2021AJ....162..198B}, where we presented the angular diameters and other fundamental stellar properties for a total of 131 stars. This paper is organized as follows: Section 2 discusses the Navy Precision Optical Interferometer and the data reduction process; Section 3 describes interferometric visibility and calibration, Section 4 details how we determined various stellar parameters, including the radius, bolometric flux, extinction; luminosity, and effective temperature for each target; Section 5 provides notes on individual stars, when required; Section 6 considers NPOI angular diameters as an ensemble; and Section 7 is the conclusion.

\section{Interferometry with the NPOI}

As mentioned previously, one of the advantages that interferometry brings is its outstanding resolution, which can be an order of magnitude better than the largest telescopes equipped with adaptive optics \citep{2020MNRAS.493.2377R}. The Navy Precision Optical Interferometer (NPOI) is located on Anderson Mesa near Flagstaff, AZ \citep{1998ApJ...496..550A, 2003AJ....125.2630H, 2003SPIE.4838..358B}. It consists of three main arms, designated north, east, and west, and incorporates two sub-arrays: the four fixed astrometric stations concentrated near the center of the array (AC, AE, AW, and AN, which stand for astrometric center, east, west, and north, respectively), and the imaging stations. The latter are labeled according to which arm they are on and their relative distance from the array center. For example, E1 is the station nearest the center on the east arm, while E10 is the station farthest away. We can combine light from the astrometric and imaging stations at will, and ``baseline'' refers to the distance between the two imaging elements. In this paper, we used 21 unique baselines, and Table \ref{baselines} lists the baselines used and their average length. The minimum length of baseline used here was just under 9 m, while the longest was just over 79 m.

For the earliest data from 1996 to 2001, we used the original version of the ``Classic'' beam combiner that recorded data on one baseline per spectrograph, of which there were three, and the light was dispersed into 32 spectral channels spanning 450 to 950 nm. The data reduction for these early years follows procedures described in \citet{1998AJ....116.2536H}. For data from 2002 on, we used the updated ``Classic'' beam combiner that records data over 16 spectral channels across 550 to 850 nm \citep{2003AJ....125.2630H, 2016ApJS..227....4H}. Every observation produced a pair of scans: a 30-second coherent (on the fringe) scan where the fringe contrast was measured every 2 ms, and an incoherent (off the fringe) scan that was used to estimate the additive bias affecting fringe measurements. 

The NPOI's data reduction package $OYSTER$ was developed by C. A. Hummel\footnote{www.eso.org/$\sim$chummel/oyster/oyster.html} and automatically edits data as described in \citet{2003AJ....125.2630H}. In addition to this process, we edited out individual data points and/or scans that showed large scatter, on the order of 5-$\sigma$ or higher. This was more common in shorter-wavelength channels where the channels are narrower, atmospheric effects are more pronounced, and the avalanche photodiode detectors have lower quantum efficiency. We removed the points because while the diameter was not affected, the error determined using these points was unfairly biased by the lower-quality shorter-wavelength channels. 

We made two assumptions about the stars at the outset: they are effectively single, and they do not rotate rapidly and therefore do not have an asymmetrical profile. Some of the targets measured here may have stellar companions, but almost all are comfortably outside of the detection sensitivity of the NPOI: \citet{2016ApJS..227....4H} showed that the NPOI can detect binaries with separations from 3 to 860 mas with magnitude differences ($\Delta m$) of 3.0 for most binary systems, and up to 3.5 when the component spectral types differ by less than two. There are a few exceptions to these assumptions, which are discussed in Section 5.

Our end goal is to obtain angular diameters on the order of 2$\%$ or less, which is considered the minimal standard of astrophysically useful measurements \citep{1997IAUS..189..147B}. Our sample consists of 30 stars with previously unpublished data in the NPOI data archive, and 3 stars observed solely in 2021, which were chosen for their large angular sizes ($\geqslant$4 mas) due to the short baselines available at the time. The dates of observations range from 1996 to 2021, and the entire data set totals more than 56,000 data points. The smallest number of measurements for a given star is 102, and the largest is 6,529. Table \ref{general_properties} includes each target's identifiers, spectral type, parallax, and metallicity ([Fe/H]), and Table \ref{observations} is the observing log.

\section{Visibility \& Calibrators}

Interferometric diameter measurements use visibility squared ($V^2$). For a point source, $V^2$ is 1 and it is defined as completely unresolved, while a star is completely resolved when its $V^2$ reaches zero. Atmospheric turbulence and instrumental effects can reduce the signal strength, significantly affecting $V^2$. In order to address this, we used calibrator stars that are small, i.e., significantly less than the resolution of the NPOI, so that $V^2$ would be at or close to 1 and is only weakly dependent on the star's angular diameter. This means we can calibrate the atmospheric and instrumental variations out of the science target measurements as we observe calibrators and science targets alternately. The observations taken during a given night were obtained using the same configuration, and the time between data collection was generally on the order of a few minutes to 10 minutes.

To estimate the calibrators' diameters, we created spectral energy distribution (SED) fits to published $UBVRIJHK$ photometry. We used plane-parallel model atmospheres from \citet{2003IAUS..210P.A20C} based on effective temperature ($T_{\rm eff}$), surface gravity (log~$g$), and $E(B-V)$. Stellar models were fit to observed photometry after converting the magnitudes to fluxes using \citet{1996AJ....112..307C} for $UBVRI$ and \citet{2003AJ....126.1090C} for $JHK$. Table \ref{calibrators} lists the photometry, $T_{\rm eff}$, log~$g$, and $E(B-V)$ used, and the resulting angular diameters.\footnote{This is a simple SED fit, unlike the more sophisticated one described in Section 4. It is an appropriate method for calibrators, given the insensitivity of the target's measured angular diameter with respect to the calibrator's diameter \citep{2018AJ....155...30B}.}

Once the visibilities are calibrated, we fit angular diameters to the data. For a uniformly-illuminated disk, $V^2 = [2 J_1(x) / x]^2$, where $J_1$ is the Bessel function of the first order, $x = \pi B \theta_{\rm UD} \lambda^{-1}$, $B$ is the projected baseline toward the star's position, $\theta_{\rm UD}$ is the apparent uniform disk angular diameter of the star, and $\lambda$ is the effective wavelength of the observation \citep{1992ARAandA..30..457S}. $\theta_{\rm UD}$ results for our program stars are listed in Table \ref{inf_results}. The data are freely available in OIFITS form \citep{2017AandA...597A...8D} upon request.

We did not stop with the uniform disk diameter, though. A more realistic model of a star's disk includes limb darkening.  When a linear limb darkening coefficient $\mu_\lambda$ is used, then
\begin{eqnarray}
V^2 = \left( {1-\mu_\lambda \over 2} + {\mu_\lambda \over 3} \right)^{-1} 
\times 
\left[(1-\mu_\lambda) {J_1(x_{\rm LD}) \over x_{\rm LD}} + \mu_\lambda {\left( \frac{\pi}{2} \right)^{1/2} \frac{J_{3/2}(x_{\rm LD})}{x_{\rm LD}^{3/2}}} \right]^2 ,
\end{eqnarray}
where $x_{\rm LD} = \pi B\theta_{\rm LD}\lambda^{-1}$ and $\theta_{\rm LD}$ is the limb darkened diameter \citep{1974MNRAS.167..475H}. We gathered published $T_{\rm eff}$, log $g$, and [Fe/H] values, and assigned a microturbulent velocity of 2 km s$^{\rm -1}$ to select $\mu_\lambda$ from \citet{2011AandA...529A..75C}. We used the ATLAS stellar model\footnote{The other choice was the PHOENIX model. We chose ATLAS because a range of metallicities were available as opposed to PHOENIX, which only had solar metallicity as an option.} in the \emph{R}-band, the waveband most closely matched to the central wavelength of the NPOI's bandpass. We note that a more refined analysis would include limb darkening's non-linear dependence on wavelength, but believe the treatment described here is fair. Limb darkening effects are related to the height of the second maximum of the visibility curve \citep{2001AandA...377..981W} and we deal almost entirely with measurements before the first minimum in this paper. 

\begin{figure}[h]
\includegraphics[width=1.0\textwidth]{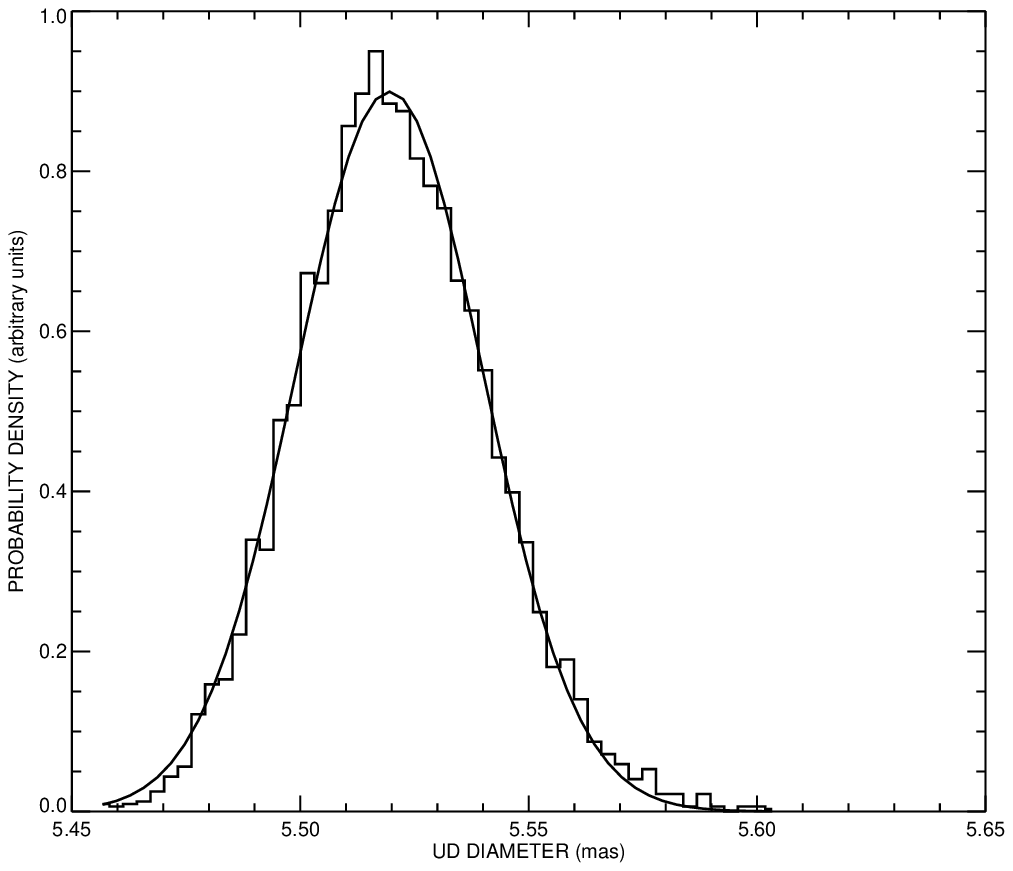}
\caption{An example probability density solution for the diameter fit to HD 3712/$\alpha$ Cas visibilities as described in Section 3.}
  \label{plot_gauss}
\end{figure}

We calculated angular diameter uncertainties using the modified bootstrap Monte Carlo method developed by \citet{2010SPIE.7734E.103T} where a large number of synthetic datasets are created by selecting entire scans at random, as opposed to a single data point within that scan. The width represents the standard deviation of the Gaussian distribution of diameters fit to these data sets, and it becomes our measure of the uncertainty for the diameter (see Figure \ref{plot_gauss}).

\begin{figure}[h]
\includegraphics[width=1.0\textwidth]{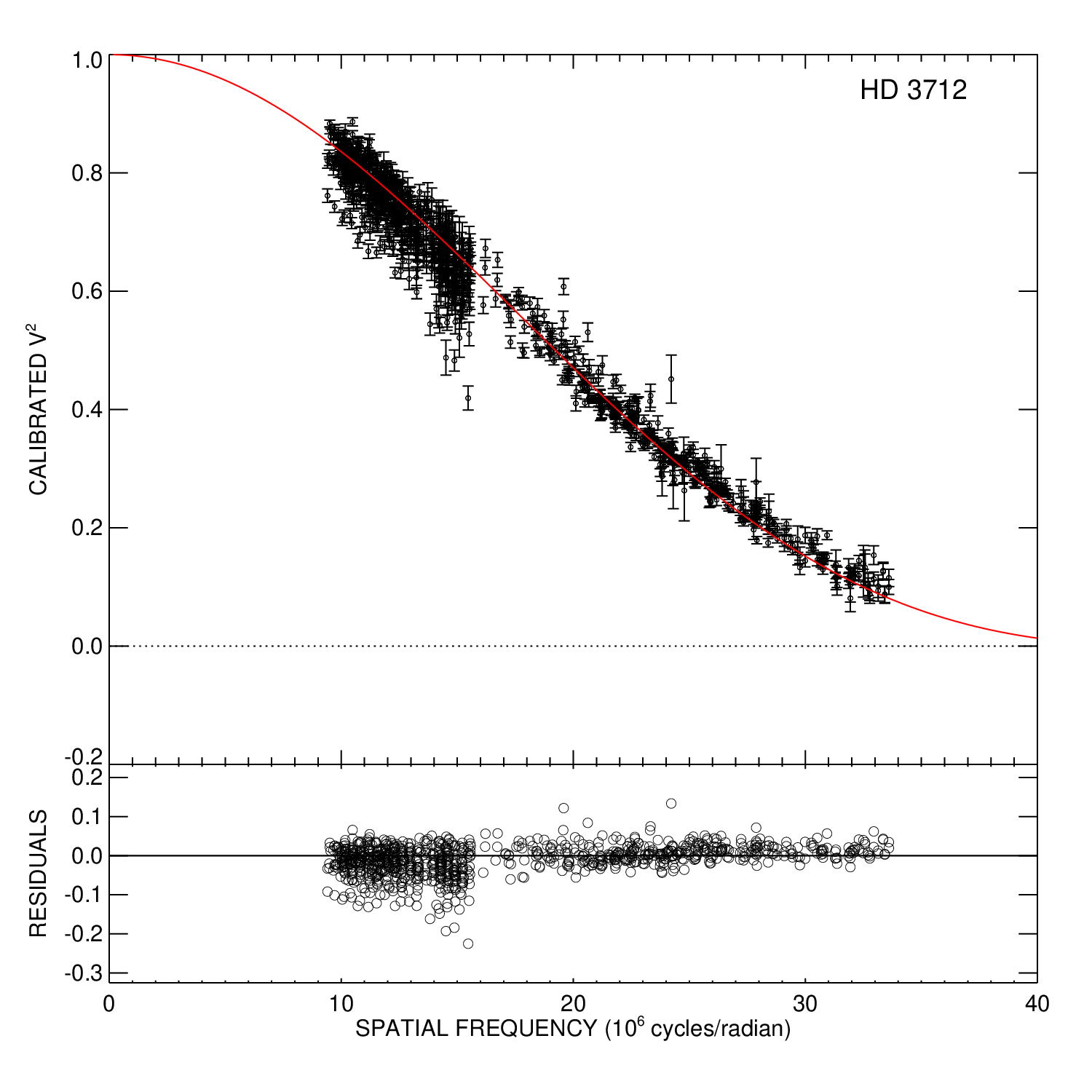}
\caption{\emph{Top panel:} The $\theta_{\rm LD}$ fit for HD 3712/$\alpha$ Cas. The solid red line represents the visibility curve for the best fit $\theta_{\rm LD}$, the points are the calibrated visibilities, and the vertical lines are the measurement uncertainties. \emph{Bottom panel:} The residuals (O-C) of the diameter fit to the visibilities. The plots for the remaining stars are available on the electronic version of the \emph{Astronomical Journal}.}
  \label{HD3712}
\end{figure}

For each target's data set, Table \ref{inf_results} shows the $T_{\rm eff}$, log $g$, [Fe/H], and $\mu_\lambda$ used, the resulting $\theta_{\rm LD}$, the maximum spatial frequency (SF), the  number of scans, and the number of data points in the angular diameter fit. Figure \ref{HD3712} shows the $\theta_{\rm LD}$ fit for HD 3712/$\alpha$ Cas as an example. The remaining plots are included in the supplementary material of the \emph{Astronomical Journal}. 

\section{Stellar Radius, Luminosity, and Effective Temperature}

When available, we converted parallax from \emph{Gaia} DR3 \citep{2022yCat.1355....0G} into a distance and combined it with our measured diameters to calculate the physical radius $R$. Otherwise, parallaxes from the \emph{Hipparcos Astrometric Catalog} \citep{2007AandA...474..653V}, \citet{2008ApJ...687.1264M}, and \emph{Gaia} DR2 \citep{2018AandA...616A...1G} were used, which was the case for 10 stars (see Table \ref{general_properties}). 

In order to determine each star's luminosity ($L$) and $T_{\rm eff}$, we created SED fits using photometric and spectrophotometric values published in the sources listed in Table \ref{phot_sources}. The assigned uncertainties for the 2MASS infrared measurements are as reported in \citet{2003yCat.2246....0C}, and an uncertainty of 0.05 mag was assigned to the optical measurements. We did not use the $R$- and $I$-band data from \citep{2002yCat.2237....0D} because they were always significant outliers.

We fit stellar spectral templates, interpolating when necessary, to the photometry from the flux-calibrated stellar spectral atlas of \citet{1998PASP..110..863P} using the $\chi^2$ minimization technique \citep{1992nrca.book.....P, 2003psa..book.....W}. This produced the bolometric flux ($F_{\rm BOL}$) and extinction ($A_{\rm V}$) for each star with the wavelength-dependent reddening relations of \citet{1989ApJ...345..245C}. Next, we combined our $F_{\rm BOL}$ values with the stars' distances ($d$) to estimate $L$ using $L = 4 \pi d^2 F_{\rm BOL}$. We also combined the $F_{\rm BOL}$ with $\theta_{\rm LD}$ to determine each star's $T_{\rm eff}$ using the equation from \citet{1999AJ....117..521V}:
\begin{equation}
F_{\rm BOL} = {1 \over 4} \theta_{\rm LD}^2 \sigma T_{\rm eff}^4,
\end{equation}
where $\sigma$ is the Stefan-Boltzmann constant and $\theta_{\rm LD}$ is in radians \citep{2014MNRAS.438.2413V}. The resulting $R$, $F_{\rm BOL}$, $A_{\rm V}$, $T_{\rm eff}$, and $L$ are listed in Table \ref{derived_results}.

Because $T_{\rm eff}$ is an input to select $\mu_\lambda$, we performed an iterative process to arrive at the final $\theta_{\rm LD}$. Table \ref{inf_results} shows the results of this process, including the initial $\theta_{\rm LD}$ and subsequent $T_{\rm eff}$, the recalculated $\mu_\lambda$, and the final $\theta_{\rm LD}$ and $T_{\rm eff}$. For six stars, $\mu_\lambda$ and $\theta_{\rm LD}$ did not change, and all of the remaining targets converged after just two iterations. Overall, $\mu_\lambda$ did not change much, with an average of 0.01 and a maximum of 0.06. The $\theta_{\rm LD}$ changed by an average of 0.4$\%$ (0.012 mas) and a maximum of 2.3$\%$ (0.055 mas). Similarly, $T_{\rm eff}$ changed an average of 9 K, and at most 46 K.

\begin{figure}[h]
\includegraphics[width=0.75\textwidth]{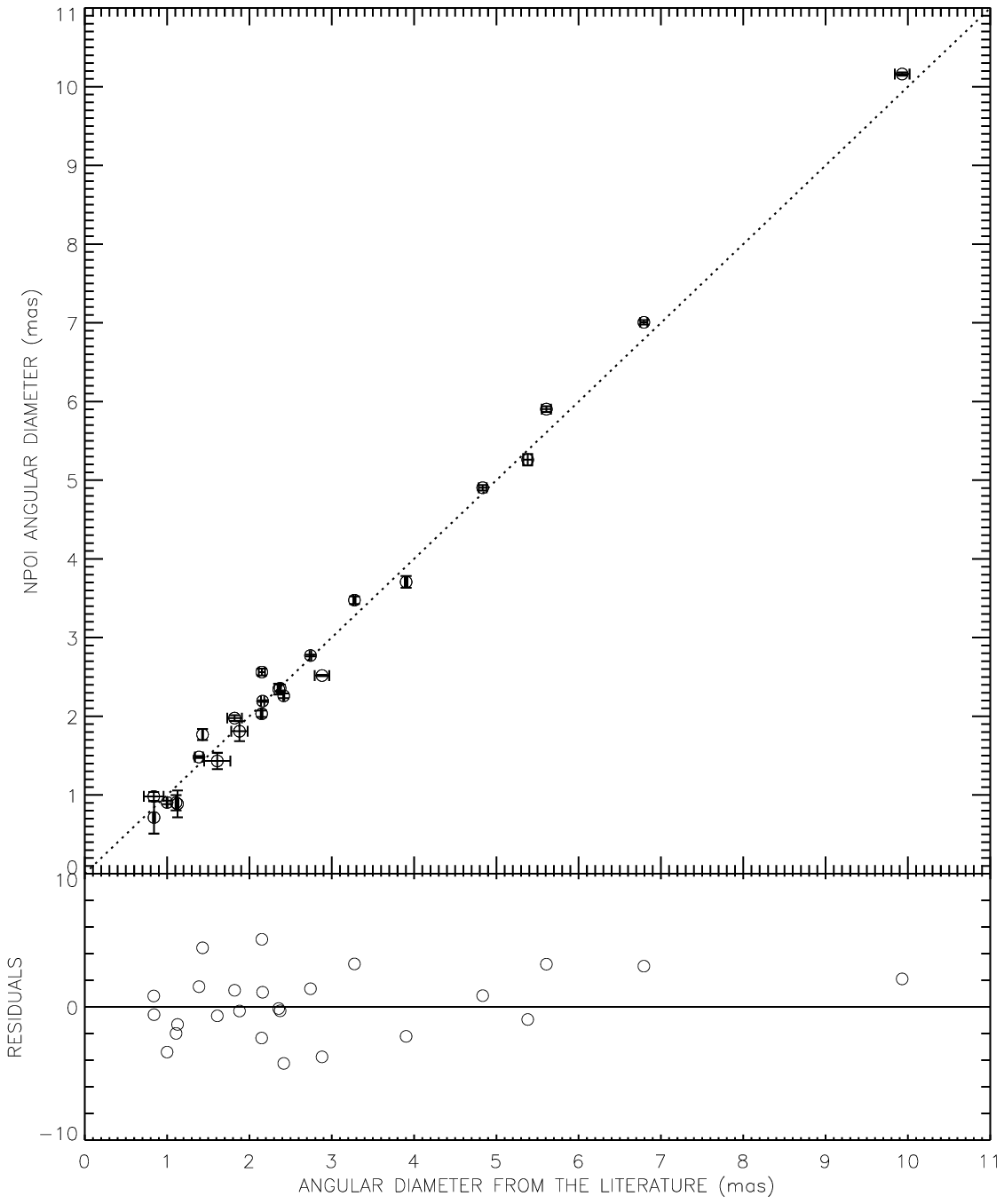}
\caption{\emph{Top panel:} Comparison of the angular diameters measured here versus previously measured interferometric diameters from the literature. The error bars are included but are often smaller than the open circle indicating the measurement. The dotted line is the 1:1 ratio. When more than one measurement was available in the literature, we used the most recent measurement (see Table \ref{lit}). \emph{Bottom panel:} The residuals were calculated as follows: ($\theta_{\rm NPOI} - \theta_{\rm literature})$ $\times$ (combined error)$^{-1}$.}
  \label{lit_diam_compare}
\end{figure}

Eight stars in this sample have never been measured before using interferometry (see Table \ref{lit}), and Figure \ref{lit_diam_compare} compares our measurements with those that came before using a variety of instruments: the Two-Telescope Stellar Interferometer at CERGA, the Mark III, the NPOI, the Center for High Angular Resolution Astronomy (CHARA) Array, the Infrared Optical Telescope Array, the Stellar Intensity Interferometer at Narrabri, the Palomar Testbed Interferometer, and the Very Large Telescope Interferometer. There is generally good agreement across instruments and the wavebands they use.

\section{Notes on Individual Stars}

Some targets of interest include the following: 

\begin{itemize}

\item \emph{HD 62044/$\sigma$ Gem:} This is a highly active single-lined spectroscopic RS CVn binary \citep{2022MNRAS.514.4190C} with imaged star spots \citep{2017ApJ...849..120R}. The companion was resolved by \citet{2015ApJ...807...23R} but the magnitude difference between the components is too large to be detected by the NPOI at $\Delta m$ = 6.72 \citep{2001AJ....122.3466M}. \citet{2017ApJ...849..120R} created temperature maps showing large star spots, though with much more tightly constrained interferometric measurements than we present here. 

\item \emph{HD 102647/$\beta$ Leo:} This young star is a favored target for exoplanet formation and evolution, considering it is one of the few known stars with hot, warm, and cold dust components with temperatures of $\sim$1600 K, 600 K, and 120 K, respectively \citep{2020ApJ...889..157C}. \citet{2021AJ....161..186D} studied $\beta$ Leo with the Large Binocular Telescope Interferometer as part of the exozodical dust survey HOSTS. They discovered a dust level some 50 times greater than the one in our solar system's zodiacal cloud, and concluded that any planet at about 5-50 AU must be less than a few Jupiter masses. They used the surface brightness relationships of \citet{2016AandA...589A.112C} to determine a $\theta_{\rm LD}$ of 1.43$\pm$0.02 mas, versus our measurement of 1.479$\pm$0.013 mas. 

In addition to the dust components, $\beta$ Leo is a $\delta$ Scuti variable and it shows pulsations, though of an unspecified type \citep{2017MNRAS.465.1181L}. It is also a multiple-star system, characterized by \citet{2015MNRAS.449.3160R} as having an A-Ba pair with a separation of 1.91\arcsec \hspace{0.01in} and $\Delta m$ = 3.9, and a Ba-Bb pair separated by 0.51\arcsec \hspace{0.01in} and $\Delta m$ = 0.129. Between the magnitude difference of the A-B pair and the fact that \citet{2009ApJ...694.1085V} considered the star a reliable star against which to compare exoplanet hosts, we treat our diameter as a single-star measurement.

\item \emph{HD 112185/$\epsilon$ UMa:} \citet{1913AN....195..369L} identified $\epsilon$ UMa as a spectroscopic binary over a hundred years ago, and \citet{2011MNRAS.413.1200R} identified a possible companion with a separation of 0.11\arcsec \hspace{0.01in} and $\Delta m$ = 2.31$\pm$0.03 in the $I-$band. We did not see any evidence of a binary companion in our data, but plan on observing the star in the future in the hope of detecting (or not) the companion. Given that this star is bright at $V$=1.77, we would expect to see a companion with that separation easily. 

\item \emph{HD 119228/83 UMa:} A suspected non-single and problem $Hipparcos$ binary \citep{1999AJ....117.1890M}, \citet{2017AJ....153..212H} observed 83 UMa with Differential Speckle Survey Instrument on the WIYN telescope, and did not detect any companions. They derived a detection limit as a function of separation and found that the $\Delta m$ at both 0.2\arcsec \hspace{0.01in} and 1.0\arcsec \hspace{0.01in} ($\Delta m$ = 4.36 and 7.55, respectively, at 692 nm, and 3.92 and 7.30, respectively, at 880 nm) would be well beyond the detection limit of the NPOI.

\item \emph{HD 124850/$\iota$ Vir:} \citet{2010ApJS..190....1R} synthesized previous information on the question of whether $\iota$ Vir is single or binary, and concluded ``single, candidate binary'' and retained it as an object for future exploration.  \citet{2012ApJ...745...24R} later used the CHARA Array to look for previously unknown companions to nearby solar-type stars to help fill the gap between spectroscopic and visual techniques. They explored the 8-80 mas range in a search for separated fringe packets, and did not find any indication of a companion for $\iota$ Vir. They again labeled it as maybe having a companion with a 55 year orbit, and we treat the star as single here.

\item \emph{HD 173764/$\beta$ Sct:} The always informative and often entertaining \citet{2008Obs...128...21G} analyzed $\beta$ Sct's binary nature and determined an orbit of P = 833.26$\pm$0.07 d. He discussed the confusion arising due to the secondary's nature, considering it is bright in ultraviolet but its contribution to the total luminosity is very small in optical wavelengths, on the order of $\Delta m$ = 4-5 in the $V$-band. \citet{2016ApJS..227....4H} used the NPOI to detect the secondary component for the first time at precisely measured separations and position angles. The $\Delta m$ of 3.6$^{\rm + 0.2}_{\rm -0.1}$ at 700 nm is on the very edge of what the NPOI can detect. 

\item \emph{HD 183912/$\beta$ Cyg A:} Part of the beautiful and increasingly complicated Albireo star system \citep[][see especially the rich collection of stars in the Washington Double Star Catalog]{2001AJ....122.3466M}, Albireo was measured by \citet{2003AJ....126.2502M} as a single star with $\theta_{\rm LD} = 4.834 \pm 0.048$ mas. More recently, \citet{2021MNRAS.502..328D} used spectroscopy to determine the all three stars in the system (Aa, Ac, and B) are likely coeval and in a hierarchical triple system with an orbital period of 121.65$^{+3.34}_{-2.90}$ years. They speculated that Alberio is not done with its surprises yet in the form of an undetected fourth companion. \citet{2022AandA...661A..49J} found evidence of that star, Ad, with a period of $\sim$ 371 d and a mass of 0.085 $M_\odot$. We did not yet search for a binary signal in the NPOI data and present the diameter as if for a single star. 

\item \emph{HD 198149/$\eta$ Cep:} \citet{2016ApJS..227....4H} obtained some tantalizing but borderline indications of binarity for this target. Of the three nights they used in their analysis, the binary model fit best for one night, marginally better for the second, and the third was better fit with a single-star model. We treat the star as single here, with our older (1997) data, and look forward to future observations and a better determination of the single or binary state of the target.

\item \emph{HD 203280/$\alpha$ Cep/Alderamin:} \citet{2006ApJ...637..494V} used the CHARA Array to characterize Alderamin as a rapid rotator with a measurable oblateness between the polar angular diameter of 1.355$\pm$0.009 mas and the equatorial diameter of 1.625$\pm$0.050 mas. This produced a rotational velocity of 283$\pm$10 km s$^{\rm -1}$, which is 83$\%$ of the breakup velocity. Also of note are the images and gravity darkening models of Alderamin made by \citet{2009ApJ...701..209Z} also made using the CHARA Array, which showed the star has two hotter, bright polar areas and a cooler, darker equator. Both sets of CHARA measurements had the advantage of better coverage around the limb of the star, more than the NPOI measurements presented here, which do not have the coverage required to detect asymmetries for this target. Our $\theta_{\rm LD}$ of 1.674$\pm$0.008 mas most likely corresponds more closely to the equatorial diameter measurement, by chance. 

\item \emph{HD 221115/70 Peg:} \citet{2009Obs...129..198G} provides the orbit for this spectroscopic binary, and notes the spectral types as G8 III for the primary and as late as M2 V for the secondary. This is beyond the sensitivity of the NPOI to detect, so we present the angular diameter as a single star.

\item \emph{HD 222107/$\lambda$ And:} A chromospherically active giant RS CVn star, \citep{2021ApJ...913...54P} imaged cool starspots using CHARA Array data from 2010 and 2011. \citet{2021ApJ...916...60M} used the same data to reconstruct temperature maps. Both teams also measured the angular diameter of the primary star, effectively single at the detection limits of interferometry, of 2.759$\pm$0.050 mas and 2.742$\pm$0.010 mas, respectively. These agree well with our measurement of 2.769$\pm$0.012 mas, though the data we used from 1997 only used three telescopes (instead of the six in the CHARA studies), so we lack the coverage to produce surface maps.

\item \emph{HD 224014/$\rho$ Cas:} This star is one of the Big Three yellow hypergiants, with the other two being HR 8752/V509 Cas and HR 5171A/V766 Cen \citep[][and references therein]{2019AandA...631A..48V}. These stars are characterized by being almost entirely convective with very extended atmospheres, having surface gravities near zero, and showing high mass-loss rates. They exhibit pulsations on quasi-periods of a few hundred days, with a pattern that shows a ``coherent sequence of pulsations, but that each pulsation is unique.'' Because of this, an SED fit would need be be better tailored to an ever-changing target, and general photometry from the literature is too smeared out to be precise. We include the angular diameter and radius in Tables \ref{inf_results} and \ref{derived_results}, but do not continue with the $T_{\rm eff}$ calculation. We also note that the phase may affect our diameter measurement, if the pulsation amplitude is significantly larger than the random error.

We used the parallax from the earlier $Gaia$ release \citep[0.947$\pm$0.202 mas, ][]{2018AandA...616A...1G} because using the more recent value from \citet{2022yCat.1355....0G} of $-0.057 \pm 0.0945$ mas produced a radius of over 4,500 $R_\odot$, and we do not claim that as real.

\end{itemize}

\section{Past and Present NPOI Diameters}

\begin{figure}[h]
\includegraphics[width=0.85\textwidth]{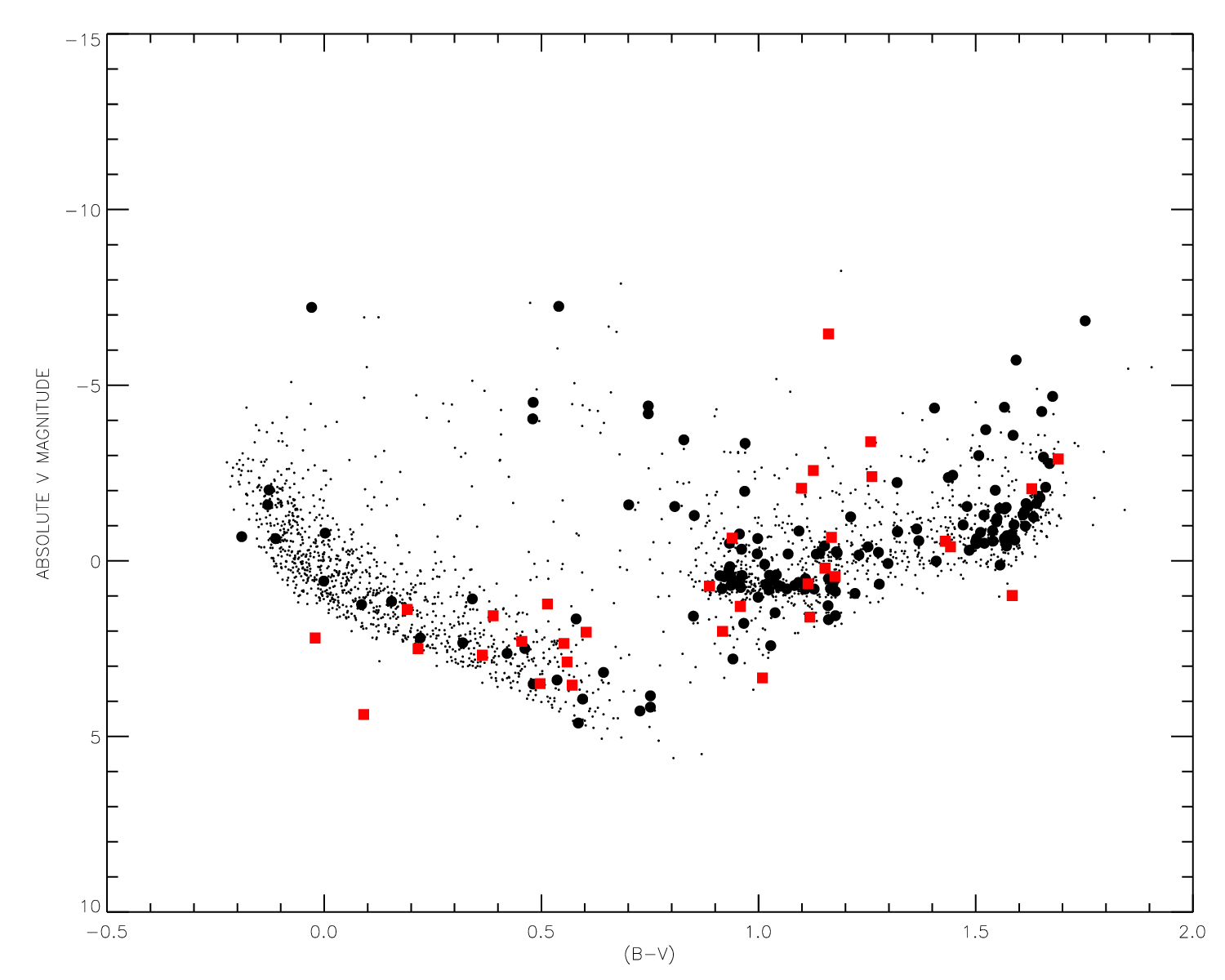}
\caption{A color-magnitude diagram of our new stars (red squares), past NPOI targets (large black circles), and targets from $JMMC$ \citep[][small black points]{2014ASPC..485..223B} that fall within the limits of the NPOI observable range of declination higher than -10 deg and brighter than $V=6.0$.}
  \label{color_mag}
\end{figure}

When we combine these stars with NPOI anguar diameters from previous samples from \citet{2014ApJ...781...90B}, \citet{2018AJ....155...30B}, \citet{2021AJ....162..198B}, and \citet{2023AJ....165...41B}, we end up with 178 stars (see Table \ref{all_diams}). Figure \ref{color_mag} shows a color-magnitude diagram for the current and previous NPOI diameters along with targets within the NPOI observing range (declination $\geq -10$ deg and $V \leq 6.0$) from the \emph{JMMC Stellar Diameters Catalog} \citep{2014ASPC..485..223B}. With a collection on this scale, we can investigate some relationships in a more quantitative sense. 

We began with the equations from \citet{2003AJ....126.2502M} linking colors to surface brightness ($S_V$):

\begin{equation}
S_V = m_V + 5 \log (\theta_{\rm LD}),
\end{equation}

where $m_V$ is the apparent visual magnitude, and 

\begin{equation}
S_V = 2.658 + 1.385 (V - K) - 0.021 (V - K)^2.
\end{equation}

\begin{figure}[h]
\includegraphics[width=0.85\textwidth]{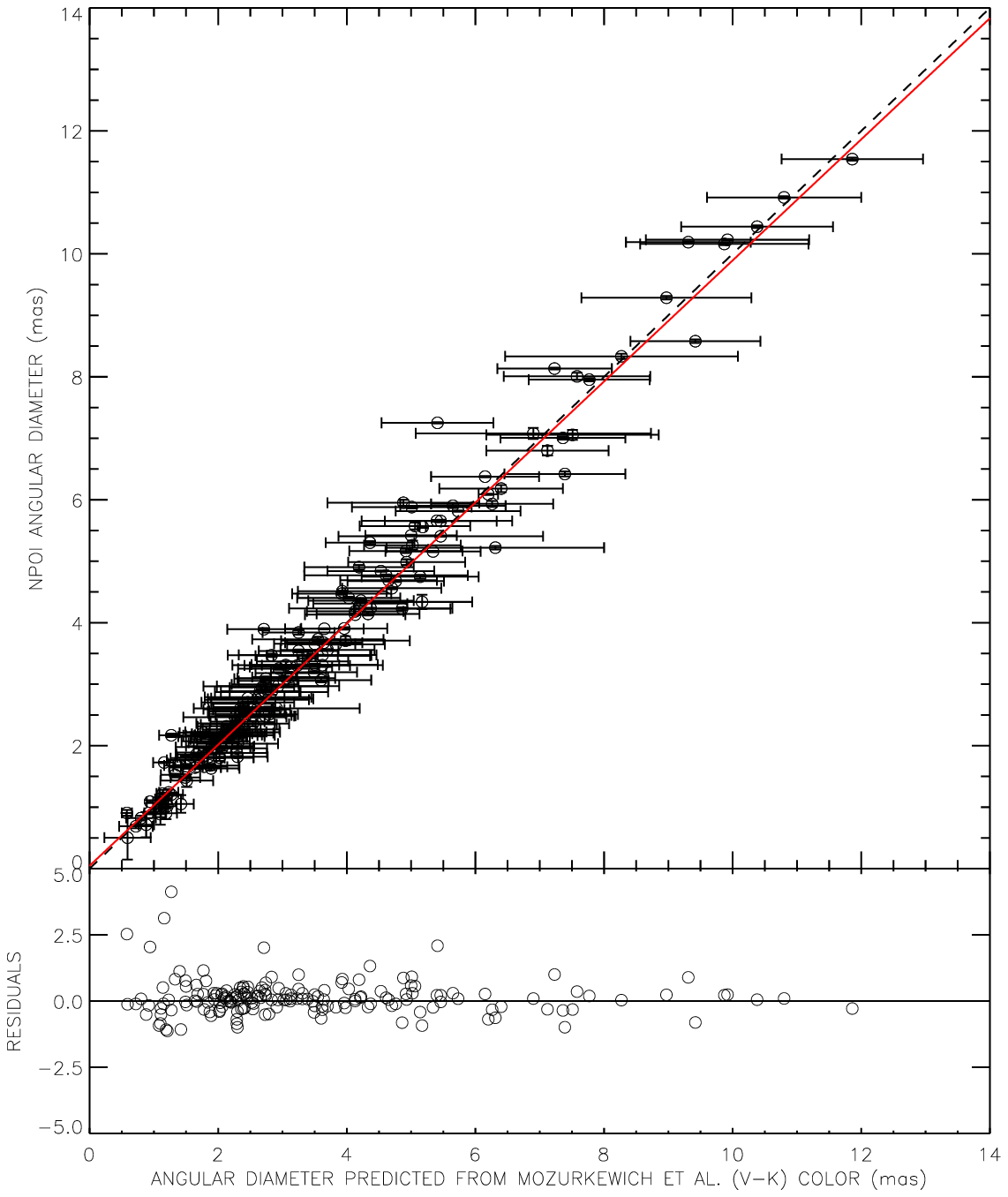}
\caption{\emph{Top panel:} Comparison of the angular diameters measured here versus diameters predicted using the relations from the \citet{2003AJ....126.2502M} paper (Eq. 3 and 4 in Section 6). Note that the NPOI errors are often smaller than the open circle indicating the data point, and are almost universally much smaller than that predicted using ($V-K$) color. The black dashed line is the 1:1 ratio, and the solid red line is the linear fit to the data ($f(x) = 1.001 x + 0.068$). \emph{Bottom panel:} The residuals to the 1:1 fit, calculated as described in Figure \ref{lit_diam_compare}.}
  \label{vk_diams}
\end{figure}

We dereddened the $V$ magnitudes from \citet{Mermilliod} using $A_{\rm V}$ listed in Table \ref{all_diams} and used Equations 3 and 4 to estimate $\theta_{\rm LD}$ from $(V-K)$ with $K$ magnitudes from \citet{2003yCat.2246....0C} except for the stars HD189319/$\gamma$ Sge and HD 192909/32 Cyg, which did not have 2MASS measurements. For these, we used $K$ magnitudes from \citet{2005AandA...434.1201R} and assigned an error of 0.01 mag because none was specified. Table \ref{all_diams} includes the resulting diameters, and Figure \ref{vk_diams} shows the result of the fit. There is excellent agreement within about 1$\%$, with a linear fit of $f(x) = 1.001 x + 0.068$.\footnote{The star HD 42995/$\eta$ Gem was removed for the fit because the $(V-K)$ diameter produced 14.07$\pm$1.02 mas while the NPOI measurement is 12.112$\pm$0.024 mas. When this star was included, the linear fit was $f(x) = 0.978 x + 0.135$. The uncertainties in the slope and intercept are $f(x)$ = 0.978$\pm$0.012 + 0.136$\pm$0.052 with HD 42995, and $f(x)$ = 1.006$\pm$0.012 + 0.069$\pm$0.050 without it.}

\begin{figure}[h]
\includegraphics[width=0.85\textwidth]{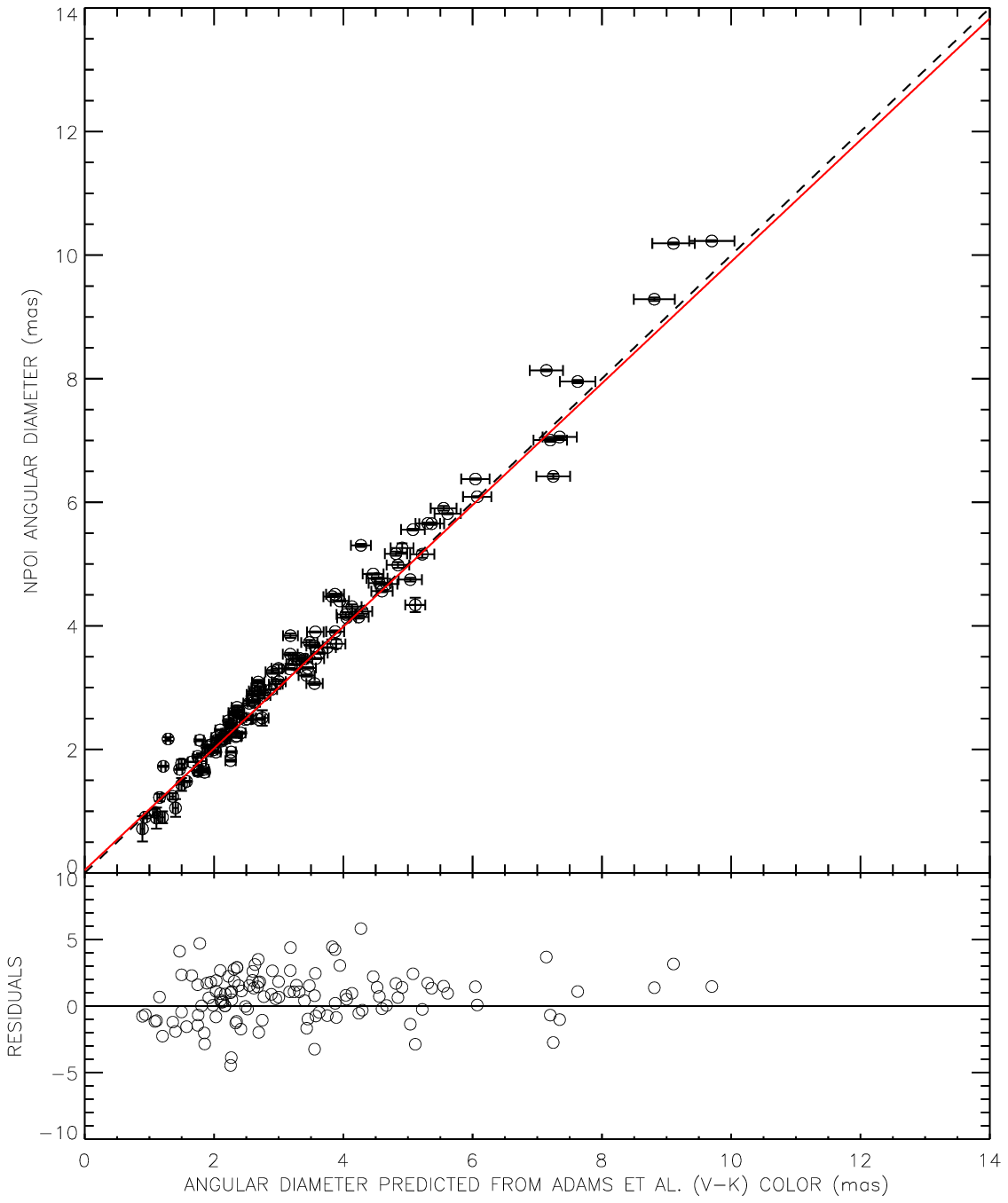}
\caption{\emph{Top panel:} Comparison of the angular diameters measured here versus diameters predicted using the relations from the \citet{2018MNRAS.473.3608A} paper. As in Figure \ref{vk_diams}, NPOI errors are often smaller than the open circle indicating the data point, as is the case for some of the diameters predicted using ($V-K$) color. The dotted line is the 1:1 ratio, and the solid red line is the linear fit to the data ($f(x) = 1.039 x - 0.014$). \emph{Bottom panel:} The residuals to the 1:1 line, calculated in the same way as described in Figure \ref{lit_diam_compare}.}
  \label{vk_diams_adams}
\end{figure}

We went through the same procedure with the angular diameter-color relations in \citet{2018MNRAS.473.3608A}, who used a sample of dwarfs/subgiants and a sample of giant stars to fit empirical relations of angular diameters to various colors, including ($V - I_C$), ($V - H$), ($V - K$), ($I_C - H$), and ($I_C - K$). Because the relations are limited to the color ranges for which they had data, we did not use all of our 178 stars: 54 of our stars were out of range while 124 were within the limits. We used the coefficients appropriate for the luminosity class of each star, and obtained a fit of $f(x) = 1.039 x - 0.014$ (see Figure \ref{vk_diams_adams}). Adams et al. provided a range of predicted fractional uncertainty, which we averaged and applied to the diameters: 3.6$\%$ for giant stars, and 3.0$\%$ for dwarf and subgiant stars.

\begin{figure}[h]
\includegraphics[width=0.75\textwidth]{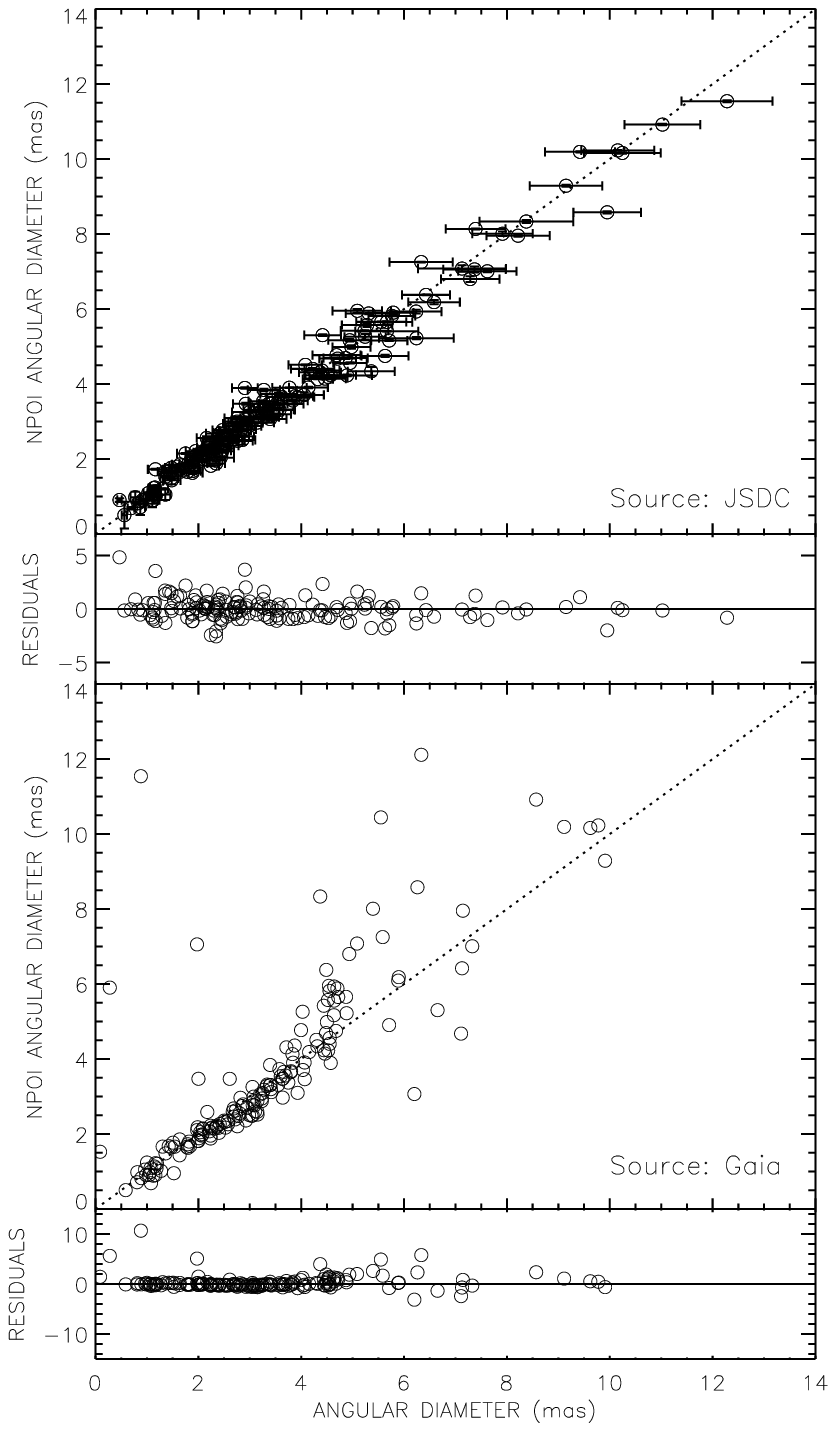}
\caption{\emph{Top panel:} Comparison of the limb-darkened angular diameters measured here versus diameters from $JMMC$ \citep{2014ASPC..485..223B}. The linear fit is $f(x) = 0.969 x + 0.088$. \emph{Bottom panel:} The same, but from $Gaia$ \citep{2018AandA...616A...1G,2019MNRAS.490.3158C}, with a linear fit of $f(x) = 0.996 x + 0.246$. Note that no errors were indicated for the $Gaia$ diameters, so the residuals are simply $\theta_{\rm NPOI} - \theta_{\rm GAIA}$.}
  \label{jmmc_gaia_diams}
\end{figure}

We also compared angular diameters from the \emph{JMMC Stellar Diameters Catalogue} \citep{2014ASPC..485..223B}, and estimates from the \emph{Gaia} catalog (Cruzal{\`e}bes et al. 2013, derived from radii and distances from \emph{Gaia} Collaboration et al. 2018). Figure \ref{jmmc_gaia_diams} shows this in graphical form, with the $JMMC$ diameters compared in the top panel and the $Gaia$ diameters in the bottom panel. The $JMMC$ diameters show a reasonable fit overall, with a linear fit of $f(x) = 0.969 x + 0.088$, while the $Gaia$ comparison shows more scatter. The fit is good with a larger y-intercept at $f(x) = 0.996 x + 0.246$, so the information could be useful for ensembles of stars, but not for an individual comparison. Cruzal{\`e}bes et al. note this issue as well, explaining that the diameters were determined using only three broad-band photometric measurements, which show strong degeneracies between $T_{\rm eff}$ and extinction/reddening meaning ``strong assumptions'' are required.

\section{Conclusion}

We measured angular diameters for 33 stars from 0.715 mas to 10.144 mas. The former has an uncertainty of $\pm$0.205 mas (29$\%$), while the latter has an uncertainty of $\pm$0.020 mas (0.2$\%$). Of the 33 stars presented here, all but six targets have diameter uncertainties of $\leq$5$\%$, and all but 12 stars have uncertainties of $\leq$2$\%$. We present six stars close to 1.0 mas or smaller, which is under the formal resolution limit of the NPOI. It is therefore not surprising that those uncertainties are amongst the highest. 

We also combined diameters from four other NPOI papers containing angular diameters to assess the collection as a whole, and compared our diameters to those obtained using other methods.

\begin{acknowledgements}

The Navy Precision Optical Interferometer is a joint project of the Naval Research Laboratory and the U.S. Naval Observatory, and is funded by the Office of Naval Research and the Oceanographer of the Navy. This research has made use of the SIMBAD database and Vizier catalogue access tool, operated at CDS, Strasbourg, France. This publication made use of data products from the Two Micron All Sky Survey, which is a joint project of the University of Massachusetts and the Infrared Processing and Analysis Center/California Institute of Technology, funded by the National Aeronautics and Space Administration and the National Science Foundation. This research has made use of the Jean-Marie Mariotti Center JSDC catalogue, available at http://www.jmmc.fr/catalogue$\_$jsdc.htm.

\end{acknowledgements}

\clearpage


\begin{deluxetable}{cc}
\tablewidth{0pc}
\tablecaption{NPOI Baselines.\label{baselines}}
\tablehead{\colhead{Telescope Pair} & \colhead{Length (m)}}
\startdata
AC-AE & 18.9	\\
AC-AN & 22.8	\\
AC-AW & 22.2	\\
AC-E3 & 9.8 	\\
AC-E6 & 34.4	\\
AC-W4 & 8.8	    \\
AC-W7 & 51.3	\\
AE-AN & 34.9	\\
AE-AW & 37.5	\\
AE-E6 & 15.9	\\
AE-W4 & 26.7	\\
AE-W7 & 64.2	\\
AN-AW & 38.2	\\
AN-W7 & 66.4	\\
AW-E3 & 31.9	\\
AW-E6 & 53.3	\\
AW-W4 & 14.0	\\
AW-W7 & 29.5	\\
E3-W4 & 18.6	\\
E6-W4 & 42.5	\\
E6-W7 & 79.4	\\
\enddata
\end{deluxetable}

\clearpage


\begin{deluxetable}{cccccccrcc}
\tablewidth{0pc}
\tablecaption{Sample Star Properties.\label{general_properties}}
\tablehead{
 \colhead{}   & \colhead{}   & \colhead{}    & \colhead{Other} & \colhead{Spectral} & \colhead{$V$}   & \colhead{Parallax} & \colhead{}    & \colhead{}       & \colhead{}    \\
 \colhead{HD} & \colhead{HR} & \colhead{FK5} & \colhead{Name}  & \colhead{Type}     & \colhead{(mag)} & \colhead{(mas)}    & \colhead{Ref} & \colhead{[Fe/H]} & \colhead{Ref} \\ }
\startdata
3712   & 168  & 21   & $\alpha$ Cas        & K0 III       & 2.23 & 14.09$\pm$0.50  & 1 & -0.10 & 5	\\
12929  & 617  & 74   & $\alpha$ Ari        & K2 III       & 2.01 & 49.56$\pm$0.25  & 2 & -0.17 & 5	\\
17361  & 824  & -    & 39 Ari              & K0.5 III     & 4.51 & 19.04$\pm$0.13  & 1 &  0.11 & 5	\\
17506  & 834  & 99   & $\eta$ Per          & K3 I-II      & 3.77 & 3.27$\pm$0.19   & 1 &  0.04 & 5	\\
37160  & 1907 & -    & $\phi^{\rm 02}$ Ori & G9.5 III     & 4.08 & 28.67$\pm$0.19  & 1 & -0.54 & 5	\\
50778  & 2574 & 266  & $\theta$ CMa        & K3/4 III     & 4.07 & 12.07$\pm$0.13  & 1 & -0.27 & 5	\\
54719  & 2697 & -    & $\tau$ Gem          & K2 III       & 4.41 & 8.33$\pm$0.16   & 1 &  0.13 & 5	\\
62044  & 2973 & -    & $\sigma$ Gem        & K1 III       & 4.26 & 27.12$\pm$0.31  & 1 & -0.06 & 5	\\
\enddata
\tablecomments{Spectral types are from SIMBAD, $V$ magnitudes are from \citet{Mermilliod}, parallaxes and [Fe/H] are from the following sources: 1. \citet{2022yCat.1355....0G}; 2. \citet{2007AandA...474..653V}; 3. \citet{2018AandA...616A...1G}; 4. \citet{2008ApJ...687.1264M}; 5. \citet{2012AstL...38..331A}; 6. \citet{2020ApJ...889..157C}; and 7. \citet{2016ApJ...826..171G}.}
\end{deluxetable}

\clearpage


\begin{deluxetable}{ccllcc}
\tablewidth{0pc}
\tablecaption{Observing Log.\label{observations}}
\tablehead{
 \colhead{Target} & \colhead{Calibrator} & \colhead{Date} & \colhead{Baselines}      & \colhead{$\#$}  & \colhead{$\#$}  \\
 \colhead{HD}     & \colhead{HD}         & \colhead{(UT)} & \colhead{Used$^\dagger$} & \colhead{Scans} & \colhead{Data Points}  \\ }
\startdata
3712   & 3360   & 2021 Oct 16 & AC-W4                              & 1  & 75 \\
       &        & 2021 Oct 20 & AC-AE, AC-W4                       & 5  & 280 \\
       &        & 2021 Oct 23 & AC-AE, AC-W4                       & 3  & 84 \\       
       & 6961   & 2021 Oct 16 & AC-W4                              & 1  & 90 \\
       &        & 2021 Oct 20 & AC-AE, AC-W4                       & 5  & 280 \\  
       &        & 2021 Oct 23 & AC-AE, AC-W4                       & 3  & 84 \\
       &        & 2021 Oct 28 & AC-AE, AW-W4                       & 2  & 56 \\     
12929  & 14055  & 2021 Oct 16 & AC-AW                              & 1  & 75 \\
\enddata
\tablecomments{See Table \ref{baselines} for the baseline lengths. ``$\#$ Scans'' is the number of scans per baseline included in the fourth column. This table shows the information for several stars as an example; the full table is available on the electronic version of the \emph{Astronomical Journal}.}
\end{deluxetable}

\clearpage


\begin{deluxetable}{lccccccccccccccc}
\tabletypesize{\scriptsize}
\tablecaption{Calibrator Stars' SED Inputs and Resulting Angular Diameters. \label{calibrators}}
\tablehead{
 \colhead{ }  & \colhead{Spec} & \colhead{$U$}   & \colhead{$B$}   & \colhead{$V$}   & \colhead{$R$}   & \colhead{$I$}   & \colhead{$J$}   & \colhead{$H$}   & \colhead{$K$}   & \colhead{$T_{\rm eff}$} & \colhead{log $g$}     & \colhead{}    & \colhead{}         & \colhead{}    & \colhead{$\theta_{\rm est}$} \\
 \colhead{HD} & \colhead{Type} & \colhead{(mag)} & \colhead{(mag)} & \colhead{(mag)} & \colhead{(mag)} & \colhead{(mag)} & \colhead{(mag)} & \colhead{(mag)} & \colhead{(mag)} & \colhead{(K)}           & \colhead{(cm s$^{-2}$)} & \colhead{Ref} & \colhead{$E(B-V)$} & \colhead{Ref} & \colhead{(mas)}   }
\startdata
886    & B2 IV    & 1.75 & 2.61 & 2.83 & 2.88 & 3.06 & 3.50 & 3.64 & 3.77 & 21944 & 3.93 & 1 & 0.02 & 8  & 0.45$\pm$0.02 \\
3360   & B2 IV    & 2.62 & 3.47 & 3.66 & 3.74 & 3.92 & 4.14 & 4.25 & 4.25 & 22180 & 3.92 & 2 & 0.03 & 9  & 0.32$\pm$0.02 \\
6961   & A7 V     & 4.63 & 4.51 & 4.34 & 4.25 & 4.16 & 4.45 & 4.28 & 4.13 & 7762  & 3.80 & 3 & 0.00 & 10 & 0.58$\pm$0.03 \\
11415  & B3 V     & 2.62 & 3.22 & 3.37 & 3.40 & 3.53 & 3.86 & 3.93 & 3.96 & 14250 & 3.38 & 4 & 0.05 & 11 & 0.50$\pm$0.03 \\
14055  & A1 V     & 4.04 & 4.03 & 4.01 & 3.99 & 4.00 & 3.80 & 3.86 & 3.96 & 9333  & 4.19 & 3 & 0.02 & 9  & 0.56$\pm$0.03 \\
17573  & B8 V     & 3.16 & 3.51 & 3.61 & 3.64 & 3.73 & 3.66 & 3.80 & 3.86 & 11749 & 4.14 & 3 & 0.01 & 9  & 0.52$\pm$0.03 \\
25642  & A0 IV    & 4.23 & 4.27 & 4.29 & 4.27 & 4.30 & 4.08 & 4.15 & 4.15 & 10900 & 3.7  & 5 & 0.08 & 12 & 0.50$\pm$0.02 \\ 
35468  & B2 V     & 0.55 & 1.41 & 1.64 & 1.71 & 1.84 & 2.15 & 2.36 & 2.38 & 21380 & 3.81 & 3 & 0.02 & 11 & 0.82$\pm$0.04 \\
\enddata
\tablecomments{Spectral types are from SIMBAD; $UBV$ values are from \citet{Mermilliod}; $RI$ values are from \citet{2003AJ....125..984M}; $JHK$ values are from \citet{2003yCat.2246....0C}; $T_{\rm eff}$, log $g$, and $E(B-V)$ values are from the following sources: 1. \citet{2007astro.ph..3658P}; 2. \citet{2004ApJS..152..251V}; 3. \citet{1999AandA...352..555A}; 4. \citet{2010ApJ...722..605H}; 5. \citet{2016AandA...589A..83G}; 6. \citet{2016AandA...591A.118S}; 7. \citet{2003AandA...402..433L}; 8. \citet{2006MNRAS.371..703S}; 9. \citet{2009AandA...501..297Z}; 10. \citet{2004AJ....127.1227C}; 11. \citet{1992BICDS..40...31F}; 12. \citet{1980BICDS..19...61N}; 13. \citet{2006MNRAS.367.1478H}; 14. \citet{2009ApJ...700.1299J}; 15. \citet{2018yCat.2354....0G}; 16. \citet{2008ApJS..176..276V}; 17. \citet{2006ApJ...647..312O}; 18. \citet{2012AandA...538A.143K}; and 19. \citet{1996AandAS..117..227A}. $\theta_{\rm est}$ is the estimated angular diameter calculated using the method described in Section 3.}
\end{deluxetable}


\begin{deluxetable}{cccccccccccccc}
\tablewidth{0pc}
\tabletypesize{\scriptsize}
\tablecaption{Interferometric Results. \label{inf_results}}

\tablehead{\colhead{Target} & \colhead{$\theta_{\rm UD}$} & \colhead{$T_{\rm eff}$} & \colhead{log $g$}        & \colhead{Initial}             & \colhead{$\theta_{\rm LD,initial}$} & \colhead{Final}               & \colhead{$\theta_{\rm LD,final}$} & \colhead{$\sigma_{\rm LD}$} & \colhead{Max SF} & \colhead{$\#$} & \colhead{$\#$} \\ 
           \colhead{HD}     & \colhead{(mas)}             & \colhead{(K)}           & \colhead{(cm s$^{-2}$)}  & \colhead{$\mu_{\rm \lambda}$} & \colhead{(mas)}             & \colhead{$\mu_{\rm \lambda}$} & \colhead{(mas)}             & \colhead{($\%$)} & \colhead{(10$^6$ cycles rad$^{\rm -1}$)}   & \colhead{scans} & \colhead{pts}            }
\startdata
3712   & 5.519$\pm$0.002 & 4690 & 1.19 & 0.67 & 5.871$\pm$0.036  & 0.71 & 5.903$\pm$0.036  & 0.6  & 33.6  & 9   & 949 	\\
12929  & 6.531$\pm$0.002 & 4439 & 2.00 & 0.70 & 6.975$\pm$0.027  & 0.73 & 7.006$\pm$0.027  & 0.4  & 33.7  & 11  & 934	\\
17361  & 1.692$\pm$0.017 & 4724 & 2.40 & 0.69 & 1.810$\pm$0.127  & 0.69 & 1.810$\pm$0.127  & 7.0  & 63.8  & 6   & 244 	\\
17506  & 4.760$\pm$0.005 & 3924 & 0.26 & 0.78 & 5.260$\pm$0.073  & 0.78 & 5.260$\pm$0.073  & 1.4  & 29.3  & 5   & 520	\\
37160  & 1.948$\pm$0.001 & 4855 & 2.61 & 0.64 & 2.188$\pm$0.010  & 0.66 & 2.193$\pm$0.010  & 0.5  & 81.1  & 56  & 2799	\\
50778  & 3.693$\pm$0.006 & 4062 & 1.36 & 0.76 & 3.725$\pm$0.074  & 0.73 & 3.707$\pm$0.074  & 2.0  & 82.2  & 8   & 208	\\
54719  & 2.348$\pm$0.011 & 7799 & 1.64 & 0.71 & 2.346$\pm$0.071  & 0.71 & 2.346$\pm$0.069  & 2.9  & 117.6 & 6   & 102	\\
62044  & 2.175$\pm$0.002 & 4591 & 2.48 & 0.71 & 2.266$\pm$0.030  & 0.69 & 2.260$\pm$0.030  & 1.3  & 67.8  & 12  & 519	\\
\enddata 
\tablecomments{The initial $\mu_{\lambda}$ is based on the $T_{\rm eff}$ and log $g$ listed here, and the final $\mu_{\lambda}$ is based on the new $T_{\rm eff}$ determination (see Section 4 for more details). The $T_{\rm eff}$ and log $g$ are from \citet{2017MNRAS.471..770M} except for HD 224014, which is from \citet{2012AandA...543A.100R}. Max SF is the maximum spatial frequency for that star's diameter measurement, $\#$ scans is the total number of scans used, and $\#$ pts is the number of data points in the angular diameter fit. \\ 
$^\ast$The diameter fit for this target may not be of significant value without knowledge of the pulsation phase of the star, as described in Section 5.} 
\end{deluxetable}

\clearpage


\begin{deluxetable}{ll}
\tablewidth{0pc}
\tablecaption{Photometry $\&$ Spectrophotometry Sources. \label{phot_sources}}
\tablehead{\colhead{Reference} & \colhead{System} }
\startdata
\citet{Mendoza}               & Johnson $UBVRI$ \\
\citet{Johnson1963}           & $IJHKLMN$ \\
\citet{Kakaras}               & $UPXYZVS$ \\
\citet{1970AandAS....1..199H} & Oja \\ 
\citet{1972VA.....14...13G}   & Geneva $VBUB_{\rm 1}B_{\rm 2}V_{\rm 1}G$ \\
\citet{1975RMxAA...1..299J}   & 13 color photometry \\
\citet{1978AandAS...34..477M} & $UBVRIJKLMNH$ \\
\citet{Alexander}             & m746, m608, m683, m710 \\
\citet{Burnashev}             & Spectrophotometry \\
\citet{1987AandAS...71..413M} & Johnson $UBV$  \\
\citet{1988iras....7.....H}   & IRAS fluxes \\
\citet{Kharitonov}            & Spectrophotometry \\
\citet{1989AandAS...81..401M} & DDO    \\
\citet{Kornilov}              & $WBVR$  \\
\citet{1993cio..book.....G}   & Catalog of Infrared Observations  \\
\citet{1995yCat.2157....0S}   & Vilinus $UPXYZVS$ \\
\citet{1997yCat.3201....0A}   & Spectrophotometry \\
\citet{1998yCat.3208....0G}   & Spectrophotometry \\
\citet{1999yCat.2225....0G}   & Catalog of Infrared Observations  \\
\citet{2000AandA...355L..27H} & Tycho $B_{\rm t}V_{\rm t}$ \\
\citet{2002yCat.2237....0D}   & Johnson $UBV$, Cousins $R_{\rm c}I_{\rm c}$ \\
\citet{2003yCat.2246....0C}   & 2MASS $JHK$ \\
\citet{2004ApJS..154..673S}   & COBE Point Source Catalog \\
\citet{2007AandA...474..653V} & Hipparcos $H_{\rm p}$ \\
\enddata
\tablecomments{These are the sources used in the SED fitting procedure described in Section 4.}
\end{deluxetable}

\clearpage


\begin{deluxetable}{ccccccccc}
\tablewidth{0pc}
\tabletypesize{\scriptsize}
\tablecaption{Derived Stellar Parameters. \label{derived_results}}

\tablehead{\colhead{Target} & \colhead{Spectral}       & \colhead{$R$}         & \colhead{$\sigma_R$} & \colhead{$F_{\rm BOL}$}                      & \colhead{$A_{\rm V}$}           & \colhead{$T_{\rm eff}$} & \colhead{$\sigma_T$} & \colhead{$L$}        \\ 
           \colhead{HD}     & \colhead{Type}           & \colhead{($R_\odot$)} & \colhead{($\%$)}     & \colhead{(10$^{-6}$ erg s$^{-1}$ cm$^{-2}$)} & \colhead{(mag)}                 & \colhead{(K)}           & \colhead{($\%$)}        & \colhead{($L_\odot$)} }
\startdata
3712   & K2 III-IV   &  45.03$\pm$$^{\rm 1.57}_{\rm 1.68}$  & 3.6  & 4.66$\pm$0.022 & 0.02$\pm$0.01 & 4476$\pm$15  & 0.3  & 734.2$\pm$52.2	\\
12929  & K2.5 III-IV &  15.19$\pm$0.10                      & 0.6  & 5.98$\pm$0.004 & 0.00$\pm$0.00 & 4373$\pm$8   & 0.2  & 76.2$\pm$0.8	\\
17361  & K2.5 IV	 &  10.22$\pm$0.72                      & 7.1  & 0.56$\pm$0.002 & 0.00$\pm$0.00 & 4768$\pm$167 & 3.5  & 48.7$\pm$0.7	\\
17506  & K3.5 I-II   & 173.10$\pm$$^{\rm 9.69}_{\rm 10.80}$ & 6.2  & 2.56$\pm$0.022 & 0.47$\pm$0.01 & 4082$\pm$30  & 0.7  & 7508$\pm$864	\\
37160  & G8 III 	 &	 8.22$\pm$0.07                      & 0.8  & 0.78$\pm$0.001 & 0.00$\pm$0.00 & 4703$\pm$11  & 0.2  & 29.8$\pm$0.4	\\
50778  & K5 III-IV   &  33.01$\pm$0.75                      & 2.3  & 1.42$\pm$0.011 & 0.11$\pm$0.01 & 4196$\pm$43  & 1.0  & 304.9$\pm$7.0	\\
54719  & K3 IV	     &	30.27$\pm$$^{\rm 1.08}_{\rm 1.09}$  & 3.6  & 0.81$\pm$0.007 & 0.31$\pm$0.01 & 4583$\pm$70  & 1.5  & 364.7$\pm$14.4	\\
62044  & K2.5 IV	 &	 8.96$\pm$0.16                      & 1.8  & 0.78$\pm$0.001 & 0.00$\pm$0.00 & 4627$\pm$31  & 0.7  & 33.2$\pm$0.8	\\
\enddata
\tablecomments{The spectral types are those that provide the best SED fit as described in Section 4. The SED fits are also the source of $F_{\rm BOL}$ and $A_{\rm V}$. The other parameters are derived as described in Section 4.}
\end{deluxetable}

\clearpage


\startlongtable
\begin{deluxetable*}{cccl}
\tabletypesize{\scriptsize}
\tablecaption{Comparing NPOI Diameters to Previous Measurements. \label{lit}}
\tablehead{\colhead{Target} & \colhead{$\theta_{\rm LD,here}$} & \colhead{$\theta_{\rm LD,literature}$} & \colhead{ }      \\ 
           \colhead{HD}     & \colhead{(mas)}                  & \colhead{(mas)}                      & \colhead{Reference} } 
\startdata
3712   & 5.903$\pm$0.036 & 5.4$\pm$0.6           & \citet{1981AandA...103...28B} \\	
       &                 & 5.4$\pm$0.6           & \citet{1983AandA...120..263F} \\	
       &                 & 5.64$\pm$0.05         & \citet{1991AJ....101.2207M}	 \\	
       &                 & 5.60$\pm$0.06         & \citet{1999AJ....118.3032N}	 \\	
       &                 & 5.72$\pm$0.08         & \citet{2001AJ....122.2707N}	 \\	
       &                 & 5.65$\pm$0.08         & \citet{2001AJ....122.2707N}	 \\	
       &                 & 5.608$\pm$0.056       & \citet{2003AJ....126.2502M}	 \\	
12929  & 7.006$\pm$0.027 & 7.6$\pm$1.0           & \citet{1983AandA...120..263F} \\
\enddata
\tablecomments{$^\ast$No LD diameter was provided, so we list the UD diameter here. Figure \ref{lit_diam_compare} shows a graphical representation of this table. If more than one diameter was available in the literature, we used the most recent one when plotting the results.}
\end{deluxetable*}

\clearpage


\startlongtable
\begin{deluxetable*}{ccccccccccc}
\tabletypesize{\scriptsize}
\tablecaption{Current and Previous NPOI Diameters as an Ensemble. \label{all_diams}}
\tablehead{\colhead{Target} & \colhead{$\theta_{\rm LD,NPOI}$} & \colhead{ }   & \colhead{$V$}  & \colhead{$A_{\rm V}$} & \colhead{$K$}   & \colhead{$\theta_{\rm LD,Mozur}$} & \colhead{$\theta_{\rm LD,Adams}$} & \colhead{Star}      & \colhead{$\theta_{\rm LD,JMMC}$} & \colhead{$\theta_{\rm LD,Gaia}$}  \\ 
           \colhead{HD}     & \colhead{(mas)}                  & \colhead{Ref} & \colhead{(mag)} & \colhead{(mag)}      & \colhead{(mag)} & \colhead{(mas)}                   & \colhead{(mas)}                   & \colhead{in range?} & \colhead{(mas)}                  & \colhead{(mas)}                    } 
\startdata
432    &  2.103$\pm$0.015 & 1 & 2.27$\pm$0.01 & 0.06 &  1.45$\pm$0.20 &  1.98$\pm$0.41 &  2.04$\pm$0.07 & N &  1.967$\pm$0.245 & 1.976	\\
1013   &  4.359$\pm$0.022 & 1 & 4.80$\pm$0.01 & 0.12 &  0.72$\pm$0.21 &  4.22$\pm$0.82 &  4.16$\pm$0.15 & N &  4.404$\pm$0.342 & 3.873	\\
1522   &  3.310$\pm$0.062 & 1 & 3.54$\pm$0.02 & 0.17 &  1.03$\pm$0.27 &  3.05$\pm$0.83 &  3.00$\pm$0.11 & Y &  3.159$\pm$0.283 & 3.552	\\
3627   &  4.185$\pm$0.036 & 2 & 3.27$\pm$0.03 & 0.10 &  0.47$\pm$0.19 &  4.13$\pm$0.75 &  4.04$\pm$0.15 & Y &  4.529$\pm$0.368 & 4.154	\\
3712   &  5.903$\pm$0.036 & 3 & 2.23$\pm$0.01 & 0.14 & -0.30$\pm$0.17 &  5.65$\pm$0.82 &  5.55$\pm$0.20 & Y &  5.789$\pm$0.394 & 0.278	\\
4656   &  3.841$\pm$0.035 & 2 & 4.43$\pm$0.01 & 0.15 &  1.11$\pm$0.17 &  3.25$\pm$0.56 &  3.18$\pm$0.11 & Y &  3.273$\pm$0.314 & 3.395	\\
5112   &  3.730$\pm$0.041 & 2 & 4.76$\pm$0.01 & 0.16 &  1.02$\pm$0.30 &  3.55$\pm$1.02 &  3.48$\pm$0.13 & Y &  3.560$\pm$0.357 & 3.576	\\
6186   &  1.887$\pm$0.025 & 4 & 4.27$\pm$0.01 & 0.11 &  2.12$\pm$0.27 &  1.77$\pm$0.43 &  1.75$\pm$0.06 & Y &  1.809$\pm$0.176 & 2.009	\\
\enddata
\tablecomments{Two stars appear in the table twice, having been published in two different NPOI papers: HD 161797 and HD 187929. \\
$\theta_{\rm LD,NPOI}$ is the NPOI limb-darkened angular diameter from the following sources: 1. \citet{2018AJ....155...30B}; 2. \citet{2021AJ....162..198B}; 3. here; 4. \citet{2023AJ....165...41B}; and 5. \citet{2014ApJ...781...90B}. \\
$V$ magnitudes are from \citet{Mermilliod}. \\
$K$ magnitudes are from \citet{2003yCat.2246....0C} for all stars except HD 189319 and HD 192909, which are from \citet{2005AandA...431..773R} and have an assigned error of 0.01. \\
$A_{\rm V}$ is from \citet{2018yCat.2354....0G}, for all but 9 stars. For those, we used: \citet[][HD 10700, HD 19373, HD 61421, and HD 114710]{2020AandA...640A...2S}, \citet[][HD 31964]{1980BICDS..19...61N}, \citet[][HD 187929]{2003AandA...402..433L}, \citet[][HD 196094, HD 208816, and HD 213311]{2005AandA...430..165F}. HD 224014 had no $A_{\rm V}$ listed on Vizier. \\
$\theta_{\rm LD,Mozur}$ is the angular diameter calculated using equations from \citet{2003AJ....126.2502M}, $\theta_{\rm LD,Adams}$ is from \citet{2018MNRAS.473.3608A}, $\theta_{\rm LD,JMMC}$ is from the $JMMC$ Stellar Diameters Catalog \citep{2014ASPC..485..223B}, and $\theta_{\rm LD,Gaia}$ is from \citet{2018AandA...616A...1G}. \\
The Y/N in the ``Star in range?'' column indicates whether or not the particular star was within the color limits of \citet{2018MNRAS.473.3608A} and is included in Figure \ref{vk_diams_adams}. See Section 6 for details.}
\end{deluxetable*}

\clearpage


\clearpage

\clearpage

\clearpage

\clearpage

\clearpage

\clearpage

\clearpage

\end{document}